# Analysis of Some Semi-Detached Systems Observed by TESS

Fahri Aliçavuş[1, a)]

[1]*Canakkale Onsekiz Mart University, Faculty of Sciences and Arts, Physics Department, 17100, Canakkale, Turkey.*

[a)]Corresponding author: fahrilcvs@gmail.com

**Abstract.** Semi-Detached binary stars are very important systems for the precise determination of astrophysical parameters and the analysis of mass transfer and loss mechanisms between the components. In this study, the light curve analysis of some semi-detached systems whose light changes are obtained by the Transiting Exoplanet Survey Satellite (TESS) was performed. The astrophysical parameters of the systems were obtained. In addition, oscillational properties of the systems were investigated, and the mass transfer and the loss amount of such stars are discussed and compared with observational data.

## INTRODUCTION

Semi-detached binary systems are very remarkable objects considering the evolutionary conditions. Interactions between components are important in terms of their effects on stellar evolution and testing of models. In addition, the effects of the components on each other and if one of the components contains oscillation, such objects come into prominence in terms of the nature of oscillations, the examination and understanding of physical processes. Accurate data obtained through the recent satellite observations increase the effect of such studies.

In this study, two semi-detached binary systems RV Pic and AS Eri are discussed. These stars have been observed by TESS and they are located inside the classical instability strip. RV Pic (HD 32011, V=9m.67, P=1.812604 day) is a neglected binary system. The most detailed study of the binary was made by Mendez (1975). In the study, RV Pic was observed in UBV bands, but a complete light curve was not obtained. Furthermore, in the same study, the radial velocities of the system were measured and only the velocities of the hot component were obtained, but no measurements of the second component were found.

The other selected semi-detached binary system is AS Eri (HD 21985, V=8m.30, P=2.664152 day) it was was first discovered by Hoffmeister (1934). In the following years, a lot of studies have been done and the absolute parameters have been determined quite well (Popper (1973), Refsdal et. al. (1974), Van Hamme and Wilson (1984)). Gamarova et al. (2000) showed that the hot component showed pulsation with a period of 24.39 minutes. Subsequently, Mkrtichian et al. (2004) revealed the existence of three different frequencies and obtained the star's oscillation modes. Finally, Narusawa (2013) conducted an abundance analysis of the primary component and discussed the relationship between pulsational characteristics and metallicity.

In this study, TESS light curves analyzes were performed for RV Pic and AS Eri and absolute parameters were calculated. In addition, oscillational properties were investigated for AS Eri.

## LIGHT CURVE SOLUTIONS AND ABSOLUTE PARAMETERS

The light curves produced by the sensitive observation data of TESS were taken over the MAST portal (https://mast.stsci.edu/portal/Mashup/Clients/Mast/Portal.html) and made ready for solution. The Wilson-Devinney (WD) (Wilson and Devinney 1971) code's MOD 5 was used for the light curve solution. Interstellar reddening was taken into account when calculating the primary components temperatures using with GAIA distances and Galactic extinction from Schlafly and Douglas (2011). The temperatures of the primary components were kept constant during the analyzes. The temperatures according to Eker et. al. (2018) were used using the calculated $(B-V)_0$ values. A detailed description of the light curve solution method is described by Soydugan et al. (2016) and Kahraman Aliçavuş and Aliçavuş (2019). Possible third-light contributions were also considered in the solutions. There was no third light effect for RV Pic, and a third light contribution of about 13% was found for AS Eri. The obtain results from the light curve solution are given in Table 1. Also, the agreement of observations with the theoretical light curves obtained from the solution is shown in Figure 1.

In addition, the basic properties of the components were determined using the parameters detemined from the light curve solutions of both binary systems. The absolute parameters for AS Eri were calculated using with radial velocities measured by Popper (1973). For RV Pic, absolute parameters were estimated using the mass corresponding to the temperature of the primary component (from Eker et al. 2018), since we do not have enough radial velocity measurement. The calculated absolute parameters are given in Table 1.

**TABLE 1.** Photometric solutions of TESS light curves and absolute parameters of the RV Pic and AS Eri

| Parameter | RV Pic | AS Eri |
|---|---|---|
| P (day) | 3.97178 | 2.664148 |
| $T_1$ (K) | 8200(150) | 8750(100) |
| $T_2$ (K) | 4221(155) | 5099(108) |
| i (deg) | 84.83(14) | 81.30(6) |
| a ($R_\odot$) | 14.24(26) | 10.35(10) |
| $\Omega_1$ | 6.3821(48) | 6.4231(79) |
| $\Omega_2$ | 2.1666 | 1.9834 |
| q | 0.174(2) | 0.108* |
| Phase Shift | 0.0000(1) | -0.0037(3) |
| $l_1/(l_1+l_2)$ | 0.806(7) | 0.729(14) |
| $l_1/(l_1+l_2)$ | 0.194(7) | 0.271(14) |
| $l_3$ | - | 0.126(13) |
| e | 0 | 0 |
| w | 90 | 90 |
| $A_1$-$A_2$ | 1.00-0.50 | 1.00-0.50 |
| $g_1$-$g_2$ | 1.00-0.32 | 1.00-032 |
| $f_1$ | 34 | 31 |
| $f_2$ | 100 | 100 |
| $r_{1(mean)}$ | 0.1609(2) | 0.1582(2) |
| $r_{2(mean)}$ | 0.2404(2) | 0.2096(2) |
| $M_1$ ($M_\odot$) | 1.95(12) | 1.895(49) |
| $M_2$ ($M_\odot$) | 0.34(2) | 0.204(10) |
| $R_1$ ($R_\odot$) | 2.30(1) | 1.642(17) |
| $R_2$ ($R_\odot$) | 3.27(1) | 2.038(19) |
| $L_1$ ($L_\odot$) | 21.4(1.6) | 14.1(7) |
| $L_1$ ($L_\odot$) | 2.96(42) | 2.5(2) |
| $Logg_1$ | 4.01(3) | 4.285(14) |
| $Logg_2$ | 2.95(3) | 3.13(2) |

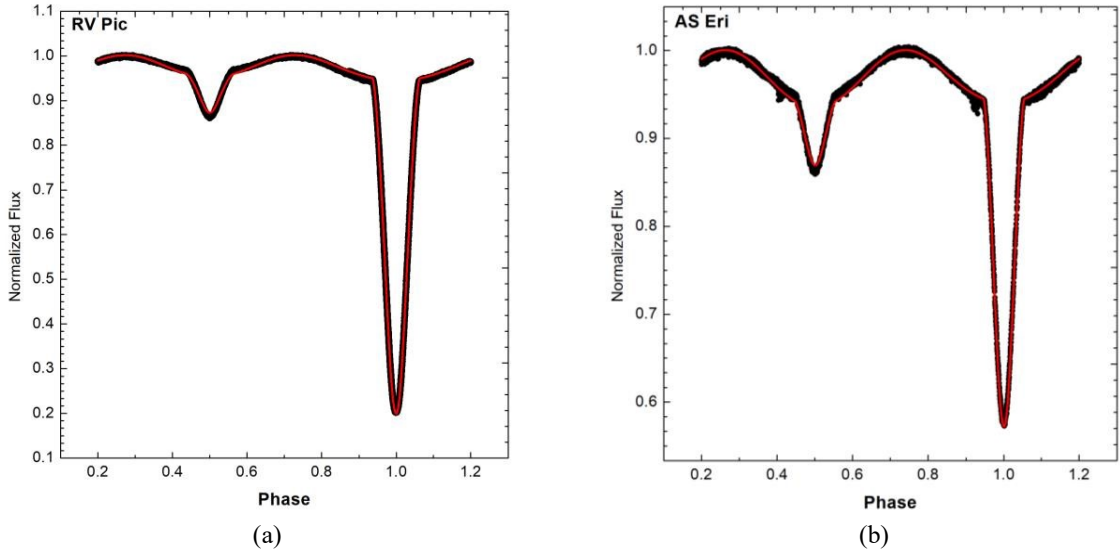

**FIGURE 1.** TESS (black) and synthetic (red) light curves of RV Pic (a) and AS Eri (b)

## PULSATION PROPERTIES

When the TESS data were analyzed, no evidence of pulsation for RV Pic was found. For AS Eri, in which pulsations were determined in previous studies, the effects caused by the orbital motion of the binary star were subtracted from the light curve and oscillational properties were examined using with TESS data. Frequency analysis was performed by using Period04 software. Significant frequency limit was taken as four sigma and analyzes were performed. As a result of the analysis, the dominant frequency was found to be 59.036 c/d which agrees with previous studies. The frequencies and amplitudes obtained from the analyzes are listed in Table 2. In addition, the model representing the frequencies and the compatibility of the observation points are given in Figure 2.

**TABLE 2.** Frequency analysis parameters for AS Eri.

|  | Frequency (c/d) | Amplitude (Flux) |
|---|---|---|
| $F_1$ | 59.036 | 0.00134 |
| $F_2$ | 65.385 | 0.00069 |
| $F_3$ | 65.156 | 0.00041 |
| $F_4$ | 58.184 | 0.00046 |
| $F_5$ | 61.937 | 0.00038 |
| $F_6$ | 61.405 | 0.00037 |
| $F_7$ | 58.795 | 0.00020 |

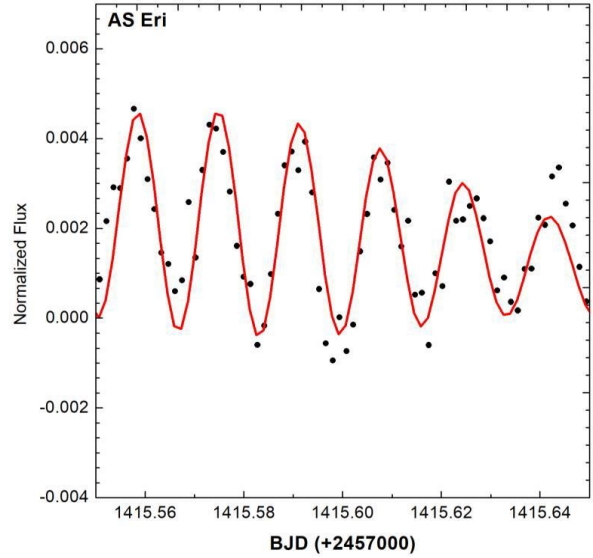

**FIGURE 2.** Out of eclipse, light variations of AS Eri. (TESS data (black) and synthetic (red) light curve)

## DISCUSSION AND CONCLUSION

In this study, TESS light curves of RV Pic and AS Eri were examined for the first time. The absolute parameters of AS Eri have been updated more sensitively. RV Pic was studied in detail for the first time and the masses of the primary and secondary components were calculated as $1.95 M_\odot$ and $0.34 M_\odot$, respectively. In addition, considering the fact that both binary stars are semi-detached, the orbital period changes were examined. However, no significant changes were observed in O-C diagrams. Considering the produced binary star evolution models, it is expected that the mass transfer and loss rates of stars with similar mass ratios will be low. The situation in the O-C diagrams in AS Eri and RV Pic indicates this.

When the pulsation frequencies of AS Eri were examined, in addition to the three frequencies Mkrtichian et. al. (2004) found, four new frequencies were found and a total of seven frequencies were determined for the hot component.

Such stars are important for understanding the evolution of stars and studying the dynamics of orbits in binary stars. Therefore, the increase of sensitive observational data is critical. Obtaining new high-resolution spectral data for RV Pic is crucial for testing our approaches to the nature of the binary system.